\documentclass[pra,twocolumn]{revtex4}
\usepackage{graphicx} 
\usepackage{amsmath} 

\begin{document}

\title{Quantum Diffusion-Limited Aggregation}

\author{David B. Johnson}
\affiliation{Department of Physics, Butler University, 4600 Sunset Ave., Indianapolis, IN 46208, USA}
\author{Gonzalo Ord\'o\~nez}
\affiliation{Department of Physics, Butler University, 4600 Sunset Ave., Indianapolis, IN 46208, USA}

\date{\today}

\begin{abstract}
Though classical random walks have been studied for many years, research concerning their quantum analogues, quantum random walks, has only come about recently. Numerous simulations of both types of walks have been run and analyzed, and are generally well-understood. Research pertaining to one of the more important properties of classical random walks, namely, their ability to build fractal structures in diffusion-limited aggregation, has been particularly noteworthy. However, only now has research begun in this area in regards to quantum random motion. The study of random walks and the structures they build has various applications in materials science. Since all processes are quantum in nature, it is important to consider the quantum variant of diffusion-limited aggregation. Recognizing that Schr\"odinger equation and a classical random walk are both diffusion equations, it is possible to connect and compare them. Using similar parameters for both equations, we ran various simulations aggregating particles. Our results show that particles moving according to Schr\"odinger equation can create fractal structures, much like the classical random walk. Furthermore, the fractal dimensions of these quantum diffusion-limited aggregates vary between 1.43 and 2, depending on the size of the initial wave packet.
\end{abstract}

\maketitle

\section{Introduction}
A Diffusion-Limited Aggregate~(DLA) is a process in which particles undergo random motion and are allowed to cluster together, forming a fractal structure~\cite{kinetic}. Computer simulations of DLA have been studied for many years leading to insights in various natural processes. For example, if the clustering property of a DLA is weakened by making aggregation less likely, the resulting structure will have a higher density. 

There can be several variations on the random motion that particles undergo in a DLA. However, it is only recently that there has been research into what happens when the rules of quantum mechanics govern the motion of a particle. Since all natural processes are truly quantum in origin, this would be an obvious next step to take in exploring the connection between DLA and nature. The objective of this study is to see how the structures generated by DLA are altered when the rules of quantum mechanics are incorporated into a DLA. Specifically, we compare the fractal dimensions of the resulting aggregates.

\begin{figure}[!ht]
\centering
\includegraphics[scale=0.25]{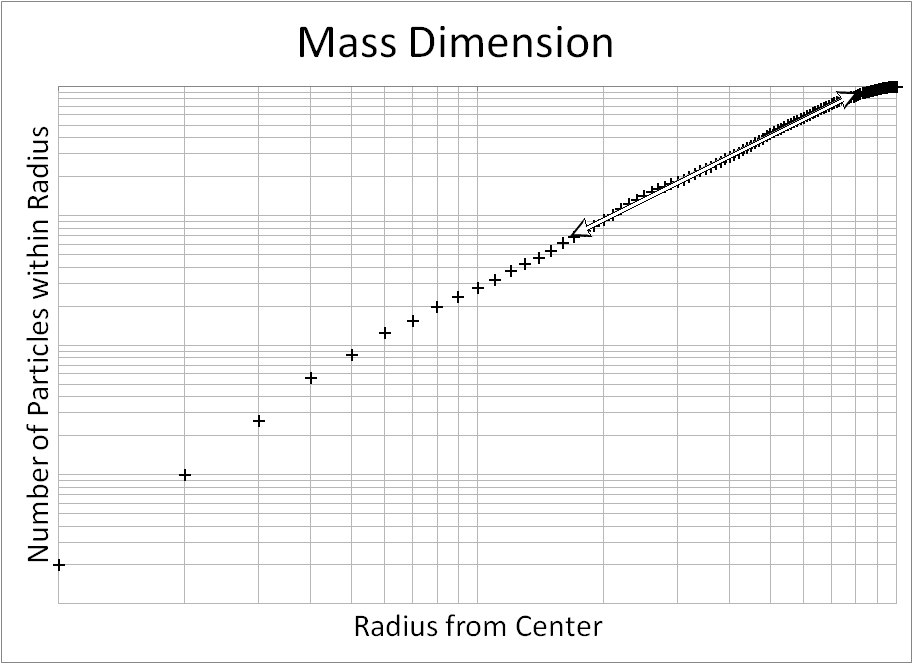}
\caption{Log-log plot of an aggregate's mass within a centered circle vs. its radius.}
\label{massdimension}
\end{figure}

Generally, the fractal dimension takes the form of a power law on some property of the fractal at different scales, where the exponent is the fractal dimension. When looking at finite structures such as those made via DLA, the fractal dimension obtained is only valid for a limited range of scales as shown in Fig.~\ref{massdimension}. In the figure, the arrowed line traces the limited range of scaling where the structure has a fractal dimension. The mass dimension assumes there is a power law relation between the radius from the center of the fractal $r$ and the mass of the fractal within that radius $M(r)$ as in Eqn.~(\ref{massdimensionequation}) where $d$ is the mass dimension and $k$ is an arbitrary constant. We use the mass dimension to measure the fractal dimension of DLA structures.

\begin{equation}
\label{massdimensionequation}
M(r) = k\,r^d
\end{equation}

Recently, there has been work on a random walk that attempts to incorporate quantum mechanics, called the quantum random walk~\cite{qrwoverview}. In a Quantum Random Walk~(QRW), the particle is in a superposition of positions instead of a single position like with the classical random walk. The probabilities amplitudes for each position are then propagated in a wave-like fashion. Previous work~\cite{colin} has shown that a quantum random walk is capable of producing fractals when used in a DLA. This work however only produced qualitative results and lacked a precise measurement of the fractal dimension of the structure formed by the quantum random walk. In this study, quantum random motion will be implemented using Schr\"odinger equation instead of a quantum random walk for reasons discussed later. 

\section{Methodology}

This study compares the structures generated in a DLA where the particles follow classical versus quantum random motion governed by Schr\"odinger's equation.

\subsection{Classical Diffusion-Limited Aggregation}

To simulate the classical random motion of the particle, two different methods have been implemented: a random walk of a particle and a diffusion of probabilities (see reference~\cite{diffusion} for a comparison btween random walks and diffusion). The equation of motion for a particle undergoing a classical random walk in two dimensions can be written as Eqn.~(\ref{crw}) where $\phi^t_{x,y}$ is the probability to find the particle at position $(x,y)$ and time $t$. As a random walk takes a step, the possible destination is evenly split between the four possible directions for a probability of one fourth in each direction. Likewise, the probability of the particle ending up in a given location is a quarter of the combined probability from all neighbor locations; this logic is captured in Eqn.~(\ref{crw}).

\begin{figure}[!ht] 
\begin{equation}
\centering
\Phi^{t+1}_{x,y}=\frac{1}{4}(\Phi^t_{x+1,y}+\Phi^t_{x-1,y}+\Phi^t_{x,y+1}+\Phi^t_{x,y-1})
\label{crw}
\end{equation}
\end{figure}

The classical random walk equation is not different from the diffusion equation~(\ref{diffusionequation}) when it is written in a numerical form~(\ref{numericaldiffusionequation}). By choosing the right parameters, the original equation for a classical random walk~(\ref{crwequation}) can be recovered from the diffusion equation. This means that the classical random walk is a diffusion process and that it can be modeled by a probability distribution via a diffusion equation~\cite[p.~44-3]{diffusion}.

\begin{figure}[!ht] 
\begin{equation}
\centering
\frac{\partial \Phi}{\partial t}=D \nabla^2 \Phi
\label{diffusionequation}
\end{equation}
\end{figure}

\begin{figure}[!ht] 
\begin{equation}
\label{numericaldiffusionequation}
\centering
\begin{split}
\frac{\Phi^{t+\Delta t}_{x,y} - \Phi^t_{x,y}}{\Delta t}=D (\frac{\Phi^t_{x+\Delta x,y}+\Phi^t_{x-\Delta x,y}-2\Phi^t_{x,y}}{(\Delta x)^2}\\
+\frac{\Phi^t_{x,y+\Delta y}+\Phi^t_{x,y-\Delta y}-2\Phi^t_{x,y} }{(\Delta y)^2})
\end{split}
\end{equation}
\end{figure}

\begin{figure}[!ht]
\begin{equation}
\label{crwequation}
\centering
\begin{split}
\Phi^{t+1}_{x,y}=\Phi^t_{x,y}+\frac{1}{4}(\Phi^t_{x+1,y}+\Phi^t_{x-1,y}+\Phi^t_{x,y+1}+\Phi^t_{x,y-1}-4\Phi^t_{x,y})\\
\Delta x=1, \Delta y=1, \Delta t=1, D=1/4 &
\end{split}
\end{equation}
\end{figure}

However, due to stability issues, it is not practical to numerically solve the diffusion equation using the parameters of equation~(\ref{crwequation}). Instead, a $\Delta t$ less than $1$ is selected, which allows for stable solutions to be numerically computed. In practice, this changes the diffusion rate $D$ also. It is assumed that this does not affect the structures generated by aggregation.

\subsection{Quantum Diffusion-Limited Aggregation}

Schr\"odinger equation~(\ref{schrodingerequation}) is also a diffusion equation, except with an imaginary diffusion coefficient,~$D$, and so can be compared to the traditional random walk~\cite{diffusion}. An explicit integration method~(\ref{explicitmethod}) is used to solve Schr\"odinger equation, where we set the potential~$V$ to zero. Whereas this scheme is unstable for real diffusion coefficients, it was selected because it becomes a stable method with the imaginary coefficient in Schr\"odinger equation~\cite{explicitintegration}. This is achieved by using a formula that is symmetrical in both space and in time, the latter of which is not the case with Eqn.~(\ref{numericaldiffusionequation}). In Eqn.~(\ref{explicitmethod}), $\psi^t_{x,y}$ is used to represent the complex-valued probability amplitude where $\psi^{t~*}_{x,y}\cdot\psi^t_{x,y}$ gives the probability to find the particle at position $(x,y)$ and time $t$.

\begin{figure}[!ht] 
\begin{equation}
\centering
i\hbar\frac{\partial \Psi}{\partial t}=-\frac{\hbar^2}{2m}(\frac{\partial^2\Psi}{\partial x^2}+\frac{\partial^2\Psi}{\partial y^2})+V(x,y)\Psi
\label{schrodingerequation}
\end{equation}
\end{figure}

\begin{figure}[!ht] 
\begin{equation}
\label{explicitmethod}
\centering
\begin{split}
i\hbar\frac{\Psi^{t+\Delta t}_{x,y}-\Psi^{t-\Delta t}_{x,y}}{2\Delta t}=-\frac{\hbar^2}{2m}(\frac{\Psi^t_{x+\Delta x,y}+\Psi^t_{x-\Delta x,y}-2\Psi^t_{x,y}}{(\Delta x)^2}\\
+\frac{\Psi^t_{x,y+\Delta y}+\Psi^t_{x,y-\Delta y}-2\Psi^t_{x,y}}{(\Delta y)^2}) &
\end{split}
\end{equation}
\end{figure}

Since Schr\"odinger equation and a classical random walk are both diffusion equations, it is possible to connect and compare them. Two programs were written: one performing a classical diffusion and the other using Schr\"odinger equation. Similar parameters were used for the quantum simulation as for the classical, when running various simulations where particles were aggregated in what we are calling Quantum Diffusion-Limited Aggregation~(QDLA). 

Pietronero et al.~\cite{fractalgrowth} have obtained a theoretical value for the fractal dimension of structures created from DLA. They considered models where particles are aggregated with a probability $P(x,y)  = \phi(x,y)^n$ where $\phi(x,y)$  obeys Laplace equation. They concluded that all such models will form a fractal for $0~\le~n~\le~2$ with a fractal dimension ranging from 2 to 1.43, respectively. Under stationary conditions, both the classical diffusion equation and Schr\"odinger equation are Laplace equations. For the classical DLA, $n$ will equal 1 while $n$ is 2 for the quantum DLA (permitting a complex $\phi(x,y)$). Therefore, we have a point of comparison between the classical and quantum DLA. 

\subsection{Implementation Details}

A square grid was created with a single point in the center, designated as the seed. Initially, a size of 256x256 was used for the grid but for later simulations, the grid was expanded to 512x512 to allow larger fractals to grow. The boundaries were set to be periodic (i.e. a torus) so that computational time was not wasted because a particle randomly leaves the grid and must be thrown away. Particles were released one at a time and allowed to run for a time period up to $T_{MAX}=500,000$ before being discarded. This value was found experimentally by allowing a free particle in an empty grid to diffuse for a long time. When the sum of the probabilities for the particle grew significantly different from unity, the accumulated error from the numerical solution to the diffusion equations was deemed too great. A fraction of this time was selected for $T_{MAX}$ to ensure the validity of the simulation. 

\begin{figure}[!ht]
\centering
\fbox{
\includegraphics[scale=0.1]{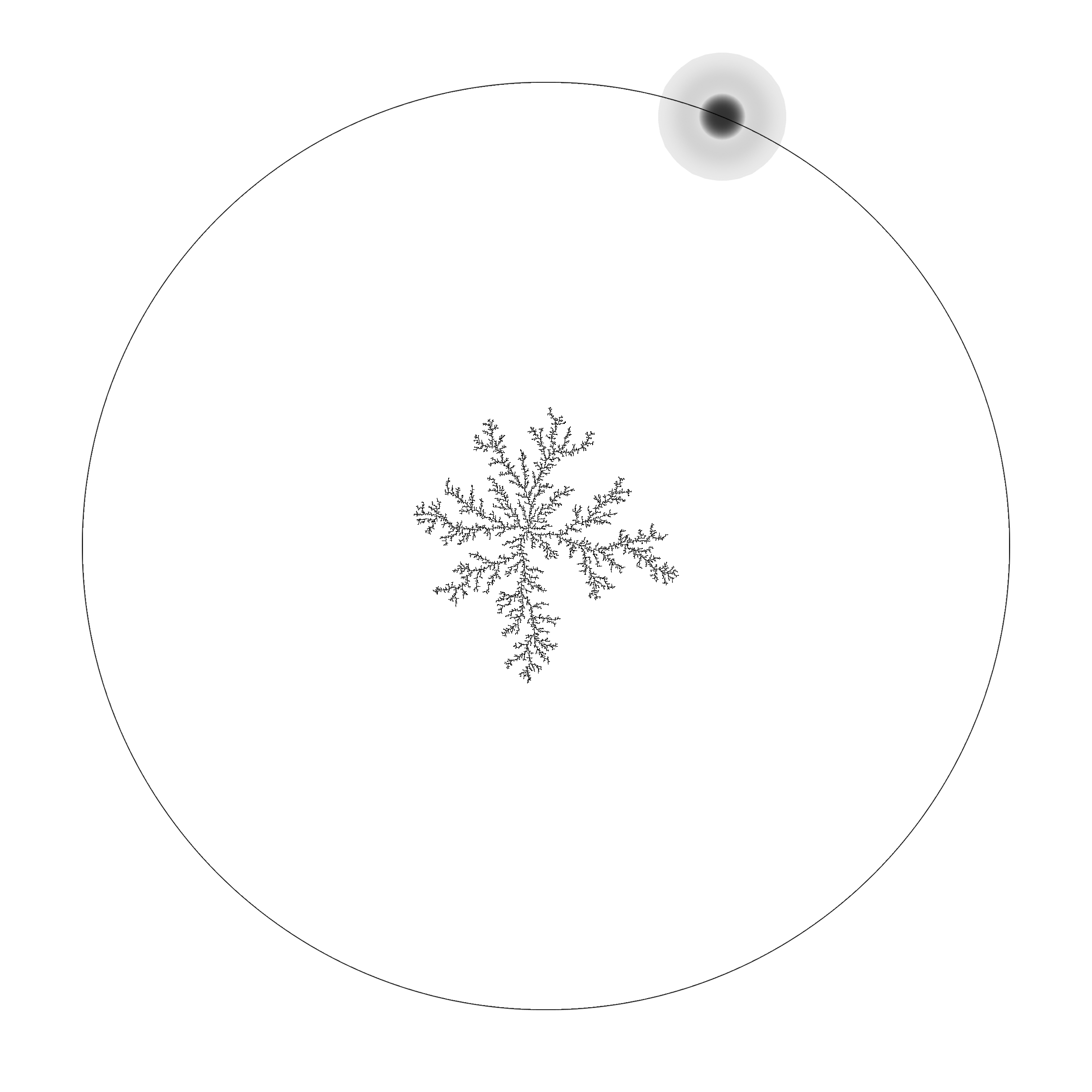}
}
\caption{The grid and a potential starting particle position.}
\label{grid}
\end{figure}

Each particle was initialized as a 2D Gaussian distribution with the standard deviations of $\sigma_x=\sigma_y=10$, which were arbitrarily selected. As with Sanberg's work~\cite{colin}, a starting distribution that is too narrow will cause the particle to interfere with itself, generating waves due to grid effects. Thus, a wider Gaussian must be selected to prevent this but it cannot be too wide because the grid has a limited size and the particle must not start out interacting with the aggregated structure. An initial velocity of zero was selected because the fractal dimension is affected the drift of the particles~\cite{drift}. This also resolves a problem in Sanberg's work because his quantum random walks had a starting bias~\cite{colin} which acts as an initial velocity~\cite{qrwoverview}. These were all issues with the original QRW-based DLA~\cite{colin} but are resolved in this study by using a 2D Gaussian distribution with no initial velocity. Every particle is placed so that it is centered randomly on the circumference of a circle, which is centered on the seed as shown in Fig.~\ref{grid}. The circumference is as wide as possible while ensuring that the particle is at least one standard deviation away from the edge. 

\begin{figure}[!ht]
\centering
\includegraphics[scale=0.5]{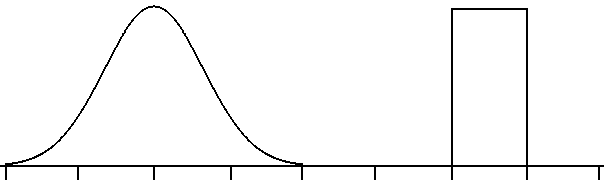}
\caption{$T=0$: Initial wave packet far from seed.}
\label{wave1}
\end{figure}

As mentioned before, the time step must be less than one~($\Delta t<1$) in order to manage the error in the numerical simulations. This requires special treatment of the propagation and detection of the particle. A 1D toy model is shown here to demonstrate the concept, which can be thought of as an exaggerated cross-section of the real simulation. When the total running time for the particle is zero~($T=0$), the probability distribution of the particle should be sufficiently far from all parts of the DLA as shown in Fig.~\ref{wave1}. 

\begin{figure}[!ht]
\centering
\includegraphics[scale=0.5]{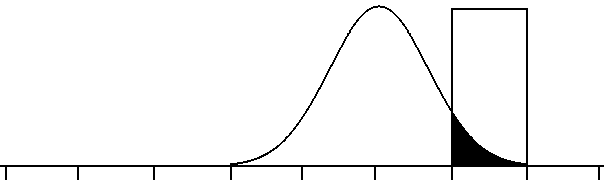}
\caption{$T=k\Delta t$: Updated wave overlapping with seed.}
\label{wave2}
\end{figure}

\begin{figure}[!ht]
\centering
\includegraphics[scale=0.5]{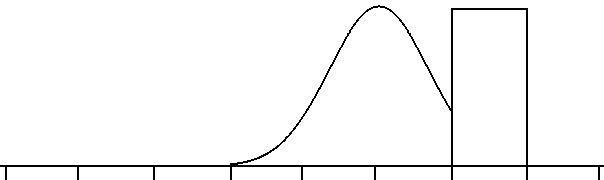}
\caption{$T=k\Delta t$: Updated wave being zeroed within seed.}
\label{wave3}
\end{figure}

As time is incremented by the time step~($T_{new}=T_{old}+\Delta t$), the quantum probability amplitudes or classical probabilities are erased at the grid locations where part of the structure is located as shown in Fig.~\ref{wave2} and Fig.~\ref{wave3}. This treats the seeds as infinite potentials where the probability of the particle entering them is zero. Consequently, the probabilities and probability amplitudes over the entire grid must be renormalized each time step.

\begin{figure}[!ht]
\centering
\includegraphics[scale=0.5]{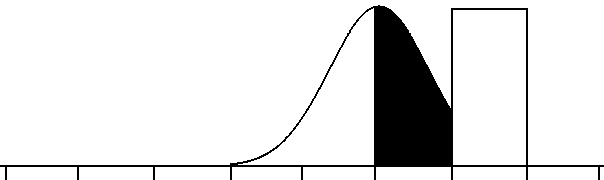}
\caption{$T=n\Delta t=1$: Wave being tested for detection.}
\label{wave4}
\end{figure}

\begin{figure}[!ht]
\centering
\includegraphics[scale=0.5]{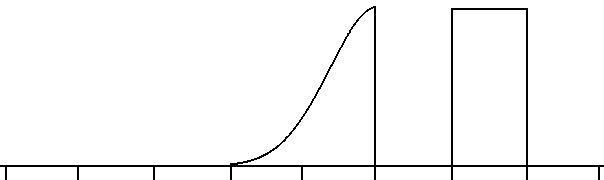}
\caption{$T=n\Delta t=1$: Wave zeroed after failing detection.}
\label{wave5}
\end{figure}

Every $n^{th}$ time step (where the time step was selected as $\Delta t = 1/n$), an attempt is made to detect the particle next to any of the seeds as in Fig.~\ref{wave4}. If it is detected, the particle is localized to that position and another particle is released. If there is no detection, all locations next to seeds have the probability amplitudes or probabilities zeroed there (requiring renormalization again) as can be seen in Fig.~\ref{wave5}. This is done because we know that the particle is definitely not at any of the locations examined.

Detection is handled the same way as in~\cite{colin}. First, a pseudorandom number is generated between 0\% and 100\%. The calculated probability of each grid location that is adjacent to a part of the DLA is added to a running total until this sum exceeds the pseudorandom number that was just generated. The grid location that causes the sum to exceed the number is where the particle is aggregated. If the total probability does not ever exceed the number generated, there is no detection. 

The simulations were all run at Butler University on the clustered supercomputer, BigDawg. The simulations were written in C, using the Message Passing Interface~(MPI) in order to leverage the parallel capabilities of BigDawg. Furthermore, multiple instances of each simulation were run in order to average the simulations and thus enhance the precision of the results. 

The program was parallelized using 16 cores with shared memory by dividing the rows of the grid between each of the cores and running the calculations in parallel. Besides needing to synchronize to ensure they remain in step together, the different cores avoided communication by relying on having concurrent read access to all needed memory. The only exception is when normalizing the wave function or performing a detection where minimal communication is necessary. 

Detection and renormalization require the sum of probabilities over the entire grid be shared between all cores. This can be done sequentially but it was parallelized in order to speed up the calculation. Every core performs the sum for its section of the grid before using a special MPI function that sums and shares the values from all cores. For grid sizes such as 512x512, it was much faster to parallelize this calculation than to have only one core perform it. If a particle is detected, a second pass over a fraction of the grid must be performed to actually determine which grid location the particle will be located. These techniques maximized parallelization and minimized communication, making the program as efficient as possible.

A utility program was written that finds the fractal dimension of a DLA. The program specifically finds the mass dimension by generating the data shown in Fig.~\ref{massdimension}. Clearly, it is not possible to just use all the data points in the graph to find the slope since the entire graph is not a straight line. Data points related to small radii suffer from grid effects, while larger radii skew the results because of the limited size of the DLA. The linear region within the curve must be identified so that its slope can be measured using least squares linear regression. Techniques developed by Kroll et al.~\cite{linearity} were used to have the program algorithmically determine the linear region instead of relying on human intuition. Then, the slope of the best fit line of the points within this linear region was used to calculate the mass dimension. 

Unfortunately, it is not a simple matter to calculate the error of the mass dimension using these techniques. Although a least squares regression allows for calculation of an error for all terms of the best fit line, there is a much larger error from selecting different points within the linear region. Therefore, providing the standard error of the slope as the error of the fractal dimension is misleading. Instead, it is better to perform numerous simulations under the exact same parameters and then present the statistics over those.

\section{Results}

\begin{figure}[!ht]
\centering
\includegraphics[scale=0.5]{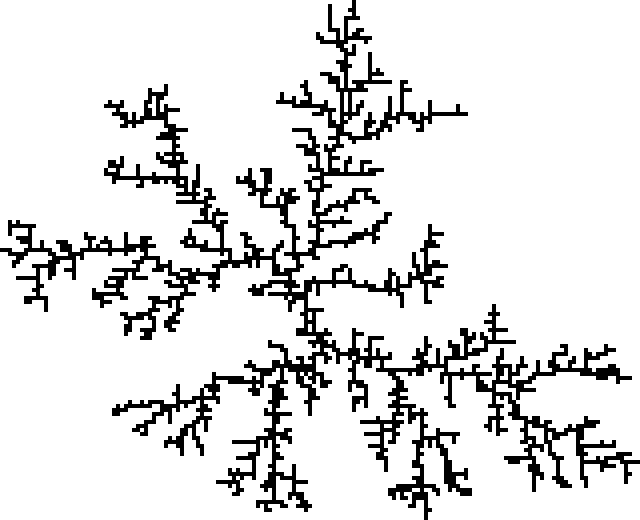}
\caption{Fractal generated by DLA using diffusion equation instead of random walk.}
\label{cdla}
\end{figure}

Because a comparison needs to be made between a classical and quantum generated DLA, this study makes the assumption that a classical random walk can be simulated as a diffusion equation without changing the resulting DLA. However, this assumption must be verified before continuing. According to Meakin~\cite{fractaldimension}, the fractal dimension of a two dimensional DLA generated via random walk is $1.69\pm0.02$. This number has been confirmed with the generation and analysis of fractals like the one in Fig.~\ref{cdla}.

\begin{figure}[!ht]
\centering
\includegraphics[scale=0.25]{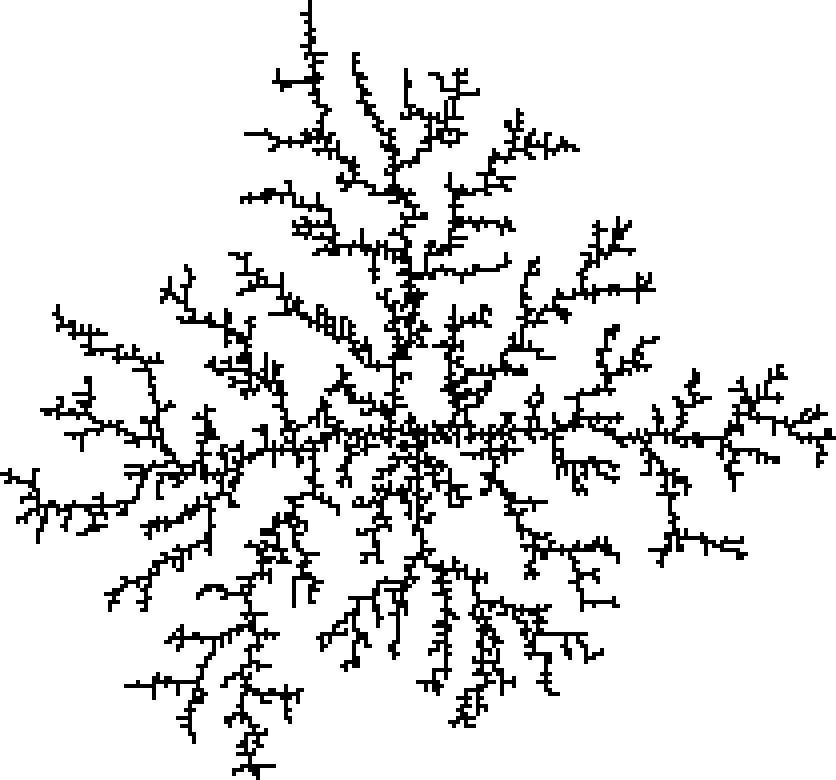}
\caption{Fractal generated by QDLA (Schr\"odinger equation).}
\label{qdla}
\end{figure}

Using identical parameters, 13 simulations of a classical DLA~(CDLA) were performed. In a CDLA, a diffusion equation is used to govern the movement of the particle instead of a random walk. This was the first patch of runs, where the grid was of the size 256x256 and the particles started randomly on a circle of radius of 113. These particles were given an initial Gaussian distribution with standard deviations $\sigma_x=\sigma_y=10$. The time step $\Delta t$ used was 0.05, which means there is an attempt at detection every $n=20$ iterations and a diffusion constant of $D=0.25$ was used. The result of these simulations is a fractal dimension of $1.67\pm0.04$, confirming that the choice of time step does not alter the generated fractals so long as the detection frequency maintains the relation $n=1/{\Delta t}$. As shown in Fig.~\ref{cdla}, it is possible to qualitatively confirm the result that fractals generated by a diffusion equation are no different than those made via random walk.

\begin{figure}[!ht]
\centering
\includegraphics[scale=0.25]{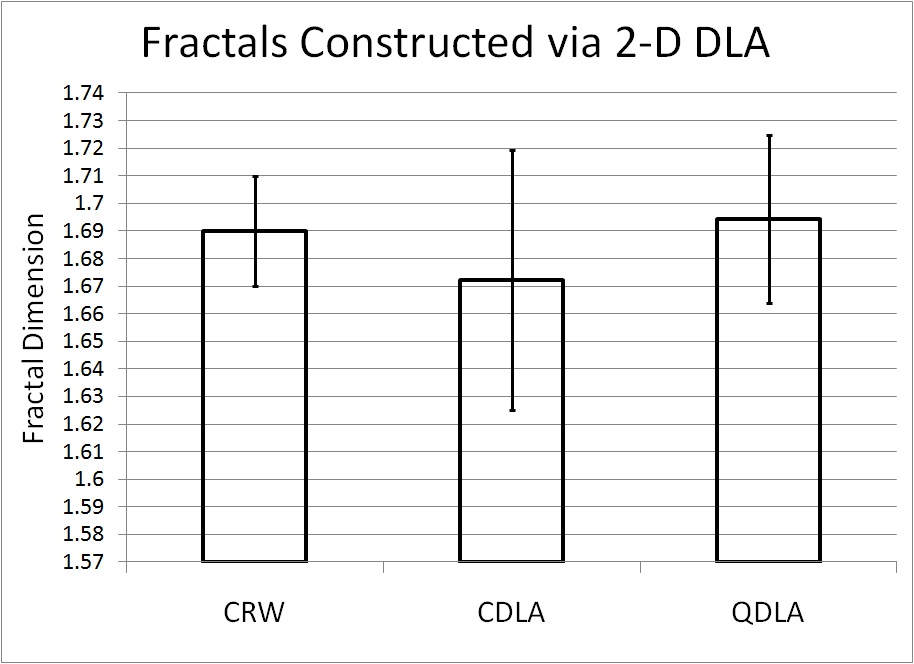}
\caption{Average fractal dimensions for several types of DLA.}
\label{dlaresults}
\end{figure}

Using the same parameters as the CDLA, a quantum DLA~(QDLA) simulation based on Schr\"odinger equation was performed in 13 identical simulations as well. In the case of the QDLA, there was an expectation for one of two possibilities. The first was that the particles would be capable of diffracting around the structure and thus will fill in the gaps between the branches of the fractal. This would lead to a fractal dimension very close to 2. The other possibility was that that diffraction does not occur and the semi-classical squared probability amplitudes would dominate, leading to a fractal dimension of 1.43 as predicted by Pietronero et al.~\cite{fractalgrowth}. From Sanberg's work~\cite{colin}, it is reasonable to expect that a fractal would be generated but the fractal dimension is unpredictable. However, we found that the average fractal dimension of the QDLA simulations was $1.69\pm0.03$ as can be visually confirmed with Fig.~\ref{qdla}. All three types of simulations generated values very close to each other as shown in Fig.~\ref{dlaresults}. This result was not expected and there is not an obvious explanation for why Schr\"odinger equation would create fractals of the same fractal dimension as a classical random walk. 

\begin{figure}[!ht]
\centering
\includegraphics[scale=0.25]{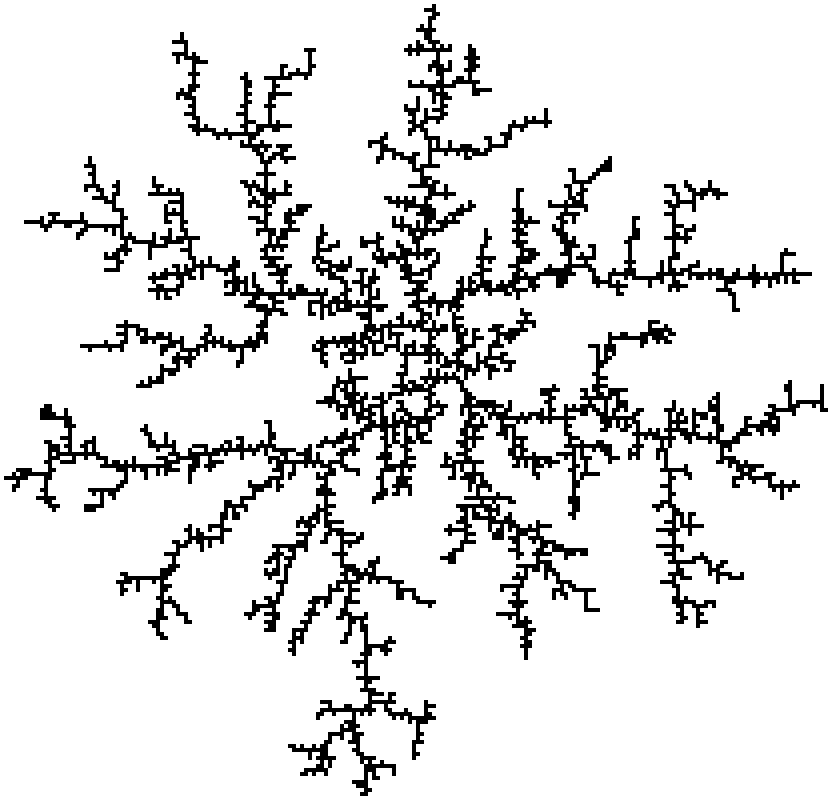}
\caption{Fractal ($d=1.45$) generated by QDLA with initial wave packet $\sigma=16$.}
\label{qdla16}
\end{figure}

\begin{figure}[!ht]
\centering
\includegraphics[scale=0.375]{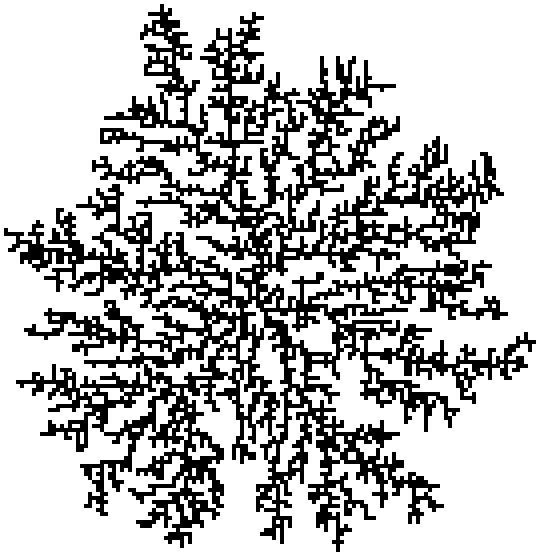}
\caption{Fractal ($d=1.91$) generated by QDLA with initial wave packet $\sigma=1$.}
\label{qdla1}
\end{figure}

Such a conclusion is very peculiar; so these results were further investigated. By examining the wave function of the QDLA interacting with the structure, it was observed that at the boundary the particle's probability amplitudes were interacting with the seeds just like how the classical diffusion equation did. The quantum particle was too spread out to be able to move between the branches. Therefore, the detections occurred in a similar fashion to the classical version. However, it was suspected that if there was a change made to the initial size of the Gaussian distribution used when initializing the particles, the particles would have different energies and thus be able to diffuse around the branches more easily. Therefore, another set of simulations was started where all of the parameters were the same but the initial wave packet size changed. Fig.~\ref{qdla16} and Fig.~\ref{qdla1} show that suddenly two very different types of fractals can result with such a change.

An additional 12 simulations were started on a 512x512 grid. Each simulation had a starting wave packet with a different size in an attempt to better characterize the relationship between the energy of the particle and the fractal dimension generated. One simulation was given a special initial configuration. There is a time invariant solution to Schr\"odinger equation in a grid with periodic boundaries such that the particle starts with equal probability everywhere. This can be thought of as equivalent to a wave packet with infinite width. This is an important configuration to consider because the particle satisfies Laplace equation when there is no seed present, which is a condition specified by Pietronero et al.~\cite{fractalgrowth}. It was expected that this run would approach the fractal dimension 1.43 that was specified.

\begin{figure}[!ht]
\centering
\includegraphics[scale=0.25]{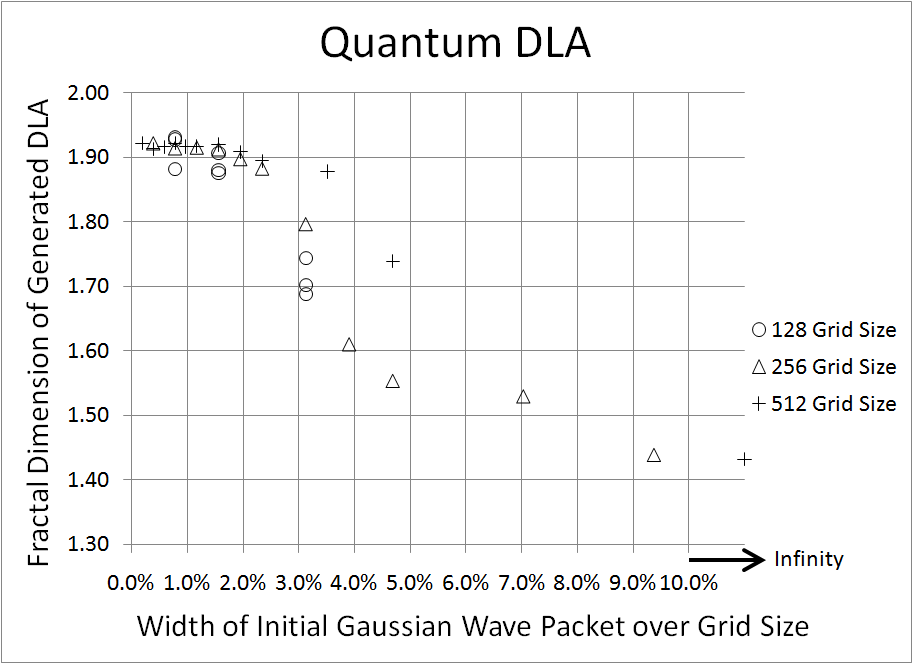}
\caption{Fractal dimension of various runs of 2D QDLA.}
\label{theresults}
\end{figure}

The fractal dimensions of all QDLA runs are shown in Fig.~\ref{theresults}. The wave packet sizes are reduced by the size of the grid so that they can be compared fairly. Unfortunately, the infinite width simulation only aggregated 768 particles after running for months. From these simulations, it was learned that the larger the wave packet, the less likely it will detect and the longer it takes to grow a DLA of significant size. So, the three 512x512 simulations with the largest wave packets should not be trusted as they did not have sufficient time to aggregate particles. Otherwise, the data seems largely consistent with some sort of curve.

\section{Conclusion} 

The data indicates that a QDLA based on Schr\"odinger equation will indeed create fractals. Furthermore, it seems that depending on the initial width of the wave packet, a fractal dimension between 1.43 and 2 can be created. It is also interesting to note that these ranges have limits that are predicted by Pietronero et al.~\cite{fractalgrowth}.

The growth of these diffusion equation based fractals was also investigated. It was found that the regions get progressively younger within the fractal as the radius increases. Typically, no detections occur between the branches because the particle's wave packet is too big and is deflected away by the tips of the branches.

For future work, it is worth taking the time to better fill in the curve of Fig.~\ref{theresults}. It is suspected that there is an inflection point where the seeping of high energy particles is in equilibrium with the deflection that occurs with low energy particles. It would be interesting to research the meaning of such a point, if it exists. It is important to not only fill in the curve but to also use an average of runs with identical starting conditions to determine the characteristic fractal dimension as well as to provide error bars. It is an interesting possibility that QDLA may be observed in experimental situations, perhaps in the deposition of particles at ultra cold temperatures.

\section*{Acknowledgements}

Very special thanks go to Dr. Dixon who imparted to us the connection between diffusion and quantum mechanics. We would like to thank Bob Holm, the Butler Institute for Research and Scholarship, and the Butler Summer Institute for funding and supporting this research.

We are extraordinarily grateful to Dr. Hatano, Dr. Petrosky, and the University of Tokyo for facilitating our trip to the International Workshop on Statistical Physics of Quantum Systems. We are also indebted to the Butler University Liberal Arts and Sciences College Deans Office, the Holcomb Undergraduate Grants Committee, Holcomb Awards Committee, and Dr. Han from Butler's Physics department for funding the trip to this conference.

This work would not be possible without the supercomputer, BigDawg, provided through funding from Dr. Levinson and the support that came with it from Nate Partenheimer. Furthermore, special thanks goes to Drs. Sorenson and Hardikar for their help in learning MPI and their support with using the supercomputer, including the willingness to give time on the supercomputer over their own programs. 

\bibliography{mybib}{}
\bibliographystyle{unsrt}

\end{document}